\documentclass[12pt]{iopart}
\usepackage{graphicx,amssymb,dsfont,bm}
\usepackage{iopams} 
\usepackage{color}
\usepackage{multirow}
\usepackage{makecell}
\usepackage{enumitem}
\usepackage{citesort}
\usepackage{longtable}
\usepackage{changepage}
\usepackage{tikz}
\usetikzlibrary{arrows.meta}
\usetikzlibrary{shadows}
\usepackage{colortbl,dcolumn}
\usepackage{tikz}
\usepackage{collcell}
\usepackage{xcolor}
\usepackage{pgf}
\usepackage{collcell}

\usepackage[title,toc,page]{appendix}
\let\appendixpagenameorig\appendixpagename
\renewcommand{\appendixpagename}{\normalsize\appendixpagenameorig}

\usepackage[titles]{tocloft}
\def\eqref#1{(\ref{#1})}

\begin{document}

\title[Statistical properties of market collective impacts]{Statistical properties of market collective responses}
\author{Shanshan Wang$^1$, Sebastian Neus\"u\ss$^2$ and Thomas Guhr$^1$}
\address{$^1$Fakult\"at f\"ur Physik, Universit\"at Duisburg--Essen, Lotharstra\ss e 1, 47048 Duisburg, Germany}
\address{$^2$Deutsche B\"orse AG, Frankfurt, Germany}
\ead{shanshan.wang@uni-due.de}
\vspace{10pt}
\begin{indented}
\item[]\today
\end{indented}

\begin{abstract}
We empirically analyze the price and liquidity responses to trade signs, traded volumes and signed traded volumes. Utilizing the singular value decomposition, we explore the interconnections of price responses and of liquidity responses across the whole market. The statistical characteristics of their singular vectors are well described by the $t$ location-scale distribution. Furthermore, we discuss the relation between prices and liquidity with respect to their overlapping factors. The factors of price and liquidity changes are non-random when these factors are related to the traded volumes. This means that the traded volumes play a critical role in the price change induced by the liquidity change. In contrast, the two kinds of factors are weakly overlapping when they are related to the trade signs and signed traded volumes. Hence, an imbalance of liquidity is related to the price change. 
\end{abstract}

\begin{adjustwidth}{2.5cm}{}
\end{adjustwidth}

\maketitle

\pagestyle{empty}
\noindent\rule{\textwidth}{1pt}
\vspace*{-1cm}
\tableofcontents
\noindent\rule{\textwidth}{1pt}
\pagestyle{headings}

\section{Introduction}
\label{sec1}

The market impact or price impact refers to the price change induced by a single trade~\cite{Bouchaud2010}. In the last two decades, it attracted ever more attention in the academic literature, as it reflects fundamental mechanisms in the market. Also from a practitioners' viewpoint, the additional transaction costs due to such price changes prompt interest in a careful data analysis. Many earlier studies on market impacts focus on single stocks~\cite{Lillo2003,Bouchaud2004,Lillo2005,Gatheral2010,Gatheral2012,Gatheral2013,Obizhaeva2013,Alfonsi2014,Alfonsi2016}. The price is determined by a continuous double auction of market orders and limit orders~\cite{Farmer2004}. Market orders are immediately executed at the available trade price, while limit orders are placed in the order book until they match another order or until they are expired or cancelled. A market order consumes the volume provided by the limit orders at the best quote, \textit{i.e.}, the best bid or the best ask. If this volume does not suffice, the market order ``goes deeper into the order book'', \textit{i.e.}, the volume of the second or even third, fourth, \textit{etc}. best quote in consumed. Hence, the best quote/price is altered, because new incoming limit orders cannot immediately supply the volumes at the previously best quote. Thus, the price change is generated due to a lack of short-run liquidity, often measured by the bid-ask spread~\cite{Demsetz1968}. The transaction cost raised in this way, \textit{i.e.}, by the self-impact, is referred to as liquidity cost~\cite{Demsetz1968}, reflecting the close connection between market impact and liquidity. 

A wealth of data made an empirical analysis of the market impact possible, and led to a microscopic understanding of the self-impact~\cite{Lillo2003,Bouchaud2004,Gabaix2003,Farmer2004}. In particular, the statistical results reveal that a long-memory correlation is present in the order flow~\cite{Bouchaud2004,Lillo2005}. This correlation is not likely to be due to herding behavior, rather it results from order splitting which is a strategy to minimize the above discussed additional transaction costs~\cite{Bouchaud2004,Toth2015}. Regarding the self-impact, many such strategies have been proposed~\cite{Gatheral2010,Gatheral2012,Gatheral2013,Obizhaeva2013,Alfonsi2014,Alfonsi2016}. 

The study of cross-impacts, \textit{i.e.}, the response of stock prices to a market order in a different stock, emerged as an obvious challenge~\cite{Hasbrouck2001,Pasquariello2013,Boulatov2013}. Large scale data analyses~\cite{Wang2016a,Wang2016b,Benzaquen2017} revealed non-Markovian features in these cross-impacts and in the corresponding trade sign cross-correlators. Consequently, the Efficient Market Hypothesis cannot hold in a strict form~\cite{Wang2016a}. Efficiency is violated on shorter time scales and only present on longer ones. There are various implications, for instance, asymmetry of information in the market~\cite{Schneider2016,Wang2017b}, latent arbitrage opportunities~\cite{Wang2017a}, or possible compensation by the cost arising from the bid-ask spread~\cite{Schneider2016}. To understand the mechanism of the cross-impacts, model-based interpretations~\cite{Wang2016c,Benzaquen2017,Patzelt2017a,Patzelt2017b} are put forward, where the liquidity once more comes into play. As limit orders provide liquidity for the market, whereas market orders take liquidity from the market, the measured liquidity results from a complex dynamical interplay. What are its characteristics? What is the role of the liquidity in the price change across the whole market? To address these issues, we apply the response function to prices and liquidity. Further, we extend the responses not only to trade signs but also to traded volumes and to the signed traded volumes. Here, we focus on the primary price and liquidity changes, which measure the price and liquidity impact without time lags, respectively.  To explore the latent factors of responses, we employ the singular value decomposition~\cite{Stewart1993,Samet2006,Sudipto2014,Benzaquen2017}. Furthermore, we discuss the relation between prices and liquidity in view of the overlapping factors.      

The paper is organized as follows. In section~\ref{sec2}, we introduce the data used in this study and the details of data processing. In section~\ref{sec3}, employing the singular value decomposition, we dissect the price and the liquidity responses to trade signs, traded volumes and signed traded volumes, respectively, and investigate the statistical properties of singular vectors. In section~\ref{sec4}, we analyze the relation between prices and liquidity in terms of their overlapping factors. We conclude our results in section~\ref{sec5}.

\section{Data description}
\label{sec2}

In section~\ref{sec2.1}, we describe the data used in this study. In section~\ref{sec2.2}, by data processing, we classify the trades into the cases of multiple trades and single trades.

\subsection{Data set}
\label{sec2.1}

The empirical analysis is carried out with the TotalView-ITCH data set, featuring 96 stocks, listed in appendix~\ref{appA}, from the NASDAQ stock market in the NASDAQ 100 index. The TotalView-ITCH data set gives detailed information about the order flow with a resolution of one millisecond. By reconstructing the order book with the order flow data, we obtain the best quote and trade information, including the best bid, the best ask, trade time, trade types (buying or selling) and so on. The details for the reconstruction of the order book are given in reference~\cite{Wang2017b}. The amount of data available for each trading day renders it possible to consider five trading days from March 7th to March 11th of 2016 for the intraday trading time from 9:40 to 15:50 in east standard time (EST) for each stock.

\subsection{Data processing}
\label{sec2.2}

\begin{figure*}[tbp]
  \begin{center}
    \includegraphics[width=0.9\textwidth]{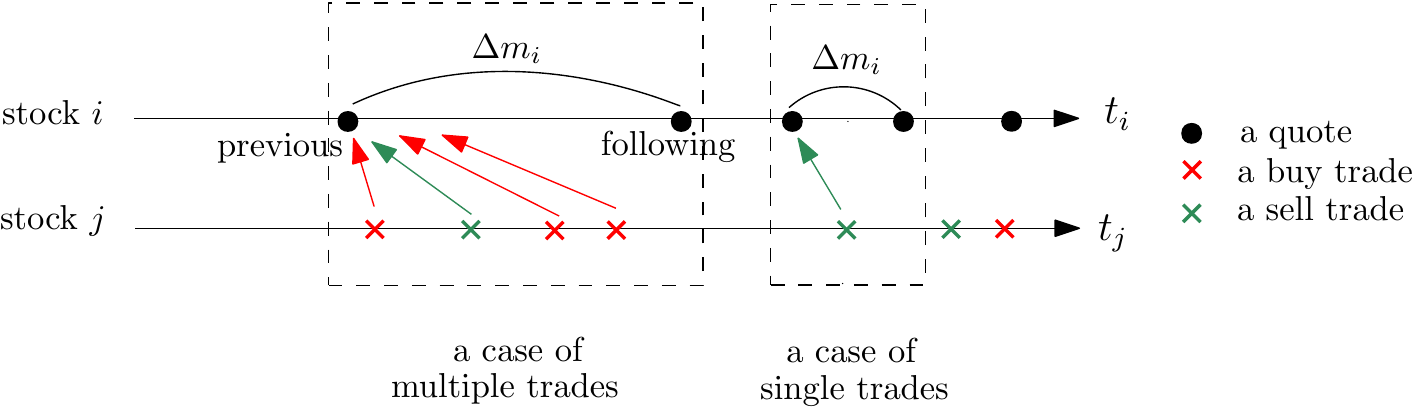}
     \caption{Sketch of data processing for the cases of multiple and single trades. The event time of stocks $i$ and $j$ are  $t_i$ and $t_j$, respectively, and $\Delta m_i$ is the price change between the previous and followed quotes of stock $i$ for each trade of stock $j$. The figure is taken from reference~\cite{Wang2017b}.}
   \label{fig2.1}
   \end{center}
   \vspace*{-0.3cm}
\end{figure*}

For each trade of a given stock, we identify the previous and following quotes of the other stock. In this way, an immediate change of quotes triggered by a trade from a different stock can be seen. The trades yield an event (trade) time axis for our study. We find that $35\%$ of the trades, on average, share their previous and following quotes with other trades, the left $65\%$ of the trades do not share their quotes with others. We refer to the former and the latter as to the cases of multiple and single trades, respectively, as sketched in figure~\ref{fig2.1}. The case of multiple trades suggests that more than one trade is needed to trigger updating the quote of the other stock.

\section{Decomposition of responses}
\label{sec3}

In section~\ref{sec3.1}, we define a generalized response function and then carry out a singular value decomposition to dissect its interconnections. We apply these methods to the price and liquidity responses in sections~\ref{sec3.2} and \ref{sec3.3}, respectively.

\subsection{Generalized response functions}
\label{sec3.1}

The average price change caused by a trade can be described by a response function~\cite{Wang2016a,Wang2016b} relating the differences of the logarithmic midpoint prices and the trade signs. The information encoded in trades not only contains trade signs for buy or sell directions, but also the traded volumes. Furthermore, the liquidity may also be changed by trades. For exploring how the price change or liquidity change is induced by trade signs, traded volumes or signed traded volumes, we generalize the response function according to  
\begin{equation}
R_{x,ij}=\left\langle \Big(\tilde{x}_i^{(f)}(t_j)-\tilde{x}_i^{(p)}(t_j) \Big) \tilde{y}_j(t_j)\right\rangle_{t_j} \ ,
 \label{eq3.1}
\end{equation}
where the indices $i$, $j$ run over all $N$ stocks, $i,j=1,\cdots, N$. The tilde above a quantity indicates the quantity normalized by
\begin{equation}
 \tilde{z}= \frac{z-\langle z\rangle}{\sigma(z)} \ ,
 \label{eq3.2}
 \end{equation}
 where $\langle z\rangle$ and $\sigma(z)$ are the mean value and the standard deviation of the corresponding time series. The normalization puts all quantities on equal footing. In equation~\eqref{eq3.1}, $\tilde{y}_j(t_j)$ stands for trade signs $\tilde{\varepsilon}_j$, traded volumes $\tilde{v}_j$, and signed traded volumes $\tilde{\nu}_j=\tilde{\varepsilon}_j\tilde{v}_j$, respectively, for stock $j$. The trade sign on an event time scale is defined either as $+1$ for a buy trade or as $-1$ for a sell trade. It can be empirically obtained from the TotalView-ITCH data set. Moreover, $\tilde{x}_i(t_j)$ in equation~\eqref{eq3.1} stands for the midpoint price $\tilde{m}_i(t_j)$ and the bid-ask spread $\tilde{s}_i(t_j)$, respectively. For $\tilde{x}_i(t_j)=\tilde{m}_i(t_j)$, equation~\eqref{eq3.1} measures the price response $R_{m,ij}$, and for $\tilde{x}_i(t_j)=\tilde{s}_i(t_j)$, equation~\eqref{eq3.1} measures the liquidity response $R_{s,ij}$. The superscript $(p)$ and $(f)$ indicate that $\tilde{x}_i(t_j)$ of stock $i$ is measured prior to or following, respectively, the trade of stock $j$ at event time $t_j$. The response function~\eqref{eq3.1} quantifies the primary price or liquidity impact of a trade, without accounting for subsequent trades. 

To explore the sources that cause the price or liquidity change in a statistical approach, we resort to a singular value decomposition of the response matrix $R_x$ for a given $x$. In our case, this matrix is a non-symmetric $N\times N$ square matrix with $N=96$, and its decomposition reads
\begin{equation}
 R_x= U_xS_xV_x^T \ 
 \label{eq3.3}
\end{equation}
where 
\begin{equation}
S_x=\textrm{diag}(s_{x1},\cdots,s_{xN})
 \label{eq3.4}
\end{equation}
is the diagonal matrix of the (real) singular values. The $N\times N$ matrices $U_x$ and $V_x$ are orthogonal, their columns $\vec{U}_{xi}$ and $\vec{V}_{xi}$, $i=1,\cdots,N$, are the left and right singular vectors, respectively, to the index $i$. We have
\begin{equation}
R_x\vec{V}_{xi}=S_{xi}\vec{U}_{xi}
\qquad
\textrm{and}
\qquad
R_x^T\vec{U}_{xi}=S_{xi}\vec{V}_{xi} \ .
\qquad
\label{eq3.5}
\end{equation}
The singular values may be identified with the latent factors that link the price or liquidity change of stock $i$ with the trading information of all other stocks $j$. The entries of the vectors $\vec{U}_{xi}$ and $\vec{V}_{xi}$ which lie between $-1$ and $+1$ are the weights of the latent factors.

According to equation~\eqref{eq3.1}, the price or liquidity change is averaged over all trades, over the trades in the case of single trades only, over the trades in the case of multiple trades only. The different averages result in the responses to all trades $R_{x,ij}|_\mathrm{at}$, to single trades $R_{x,ij}|_\mathrm{st}$, to multiple trades $R_{x,ij}|_\mathrm{mt}$. Moreover, we introduce a weight factor $w_{ij}$ to define a linearly interpolating weighted responses
\begin{equation}
R_{x,ij}|_\mathrm{wt}=w_{ij}R_{x,ij}|_\mathrm{st}+(1-w_{ij})R_{x,ij}|_\mathrm{mt} \ ,
\label{eq3.6}
\end{equation}
where $w_{ij}$ is the ratio of trades identified as single trades to all trades for a stock pair ($i$, $j$).

\subsection{Decomposing price responses}
\label{sec3.2}

Instead of looking at the single values, It is advantageous to analyze the singular vectors. One of reasons is the better statistics when the number of stocks is small. More importantly, the singular vectors disclose the correlations between price changes (trades) and the latent factors. A large correlation in the left singular vectors indicates a pronounced dependence of price change due to the corresponding factor. A large correlation in the right singular vectors implies that the factor is robustly associated with the trade. To have a overall view of these correlations in the market, we work out the statistical distribution of the entries of the singular vectors.  

\begin{figure*}[pb]
  \begin{center}
    \includegraphics[width=0.9\textwidth]{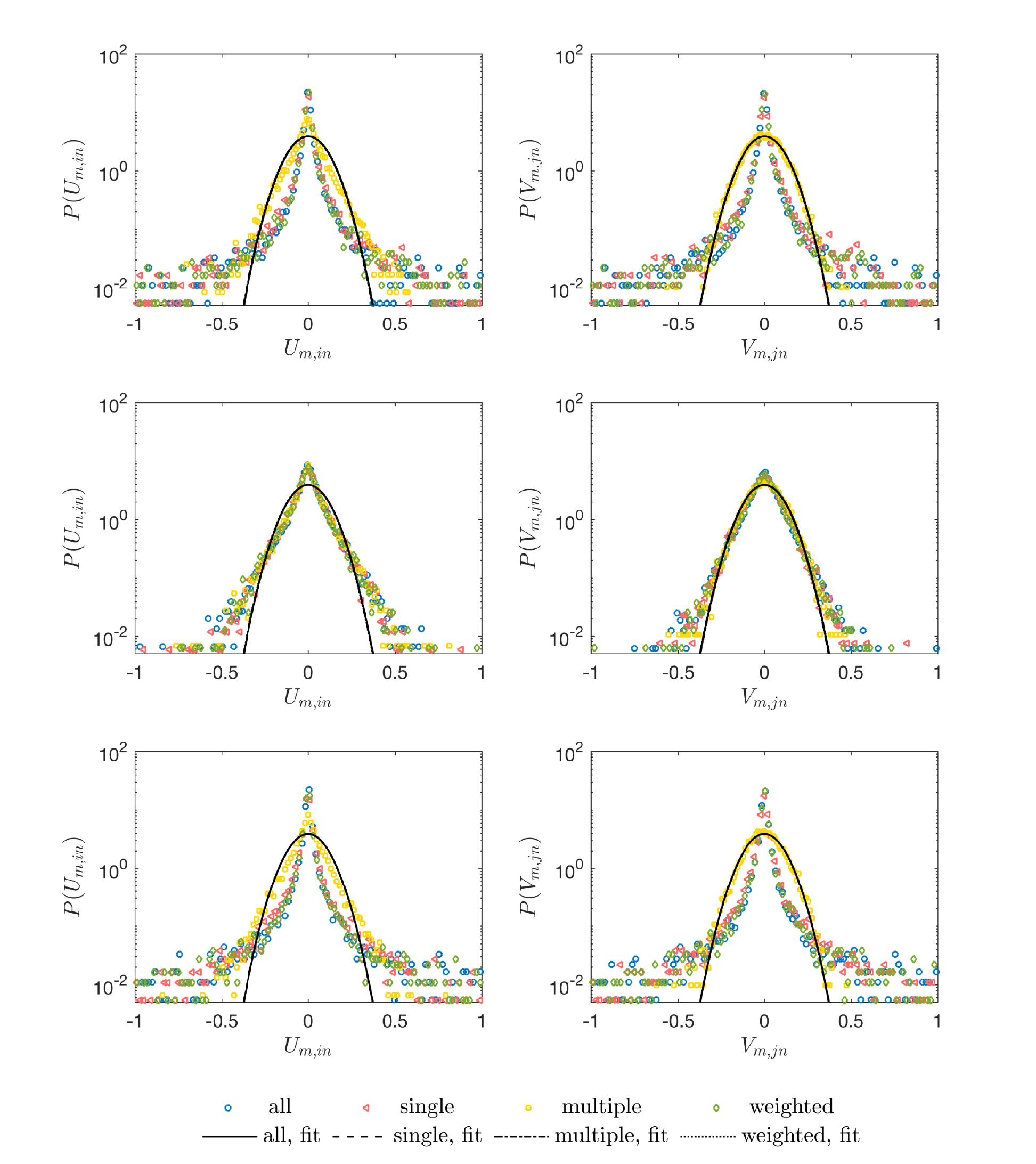} 
    \caption{Probability density distributions of the entries of left and right singular vectors $U_{m,in}$ and $V_{m,jn}$ for the price response across the whole market. Top: price response to trade signs; middle: price response to traded volumes; bottom: price response to signed traded volumes. The fits to normal distributions are shown as black lines. In each subplot, the four cases of responses, \textit{i.e.}, the case of all trades, the case of single trades, the case of multiple trades and the case of weighted trades, are displayed.}
   \label{fig3.2}
   \end{center}
\end{figure*}

Figure~\ref{fig3.2} shows the empirical probability densities of the left and right singular vectors, \textit{i.e.}, $U_{m,in}$ and $V_{m,jn}$ with the factors $n=1,\cdots, N$, for the four types of price responses across the whole market. For comparison, normal distributions are fitted to the empirical distributions. We find that the price change is not occasional and must be induced by some factors. No matter the price responds to trade signs, traded volumes or signed traded volumes, the case of multiple trades is close to random in the correlation between trades and the latent factors. On the contrary, the other cases with the similar distributions are more informative.

\begin{figure*}[pb]
  \begin{center}
    \includegraphics[width=0.9\textwidth]{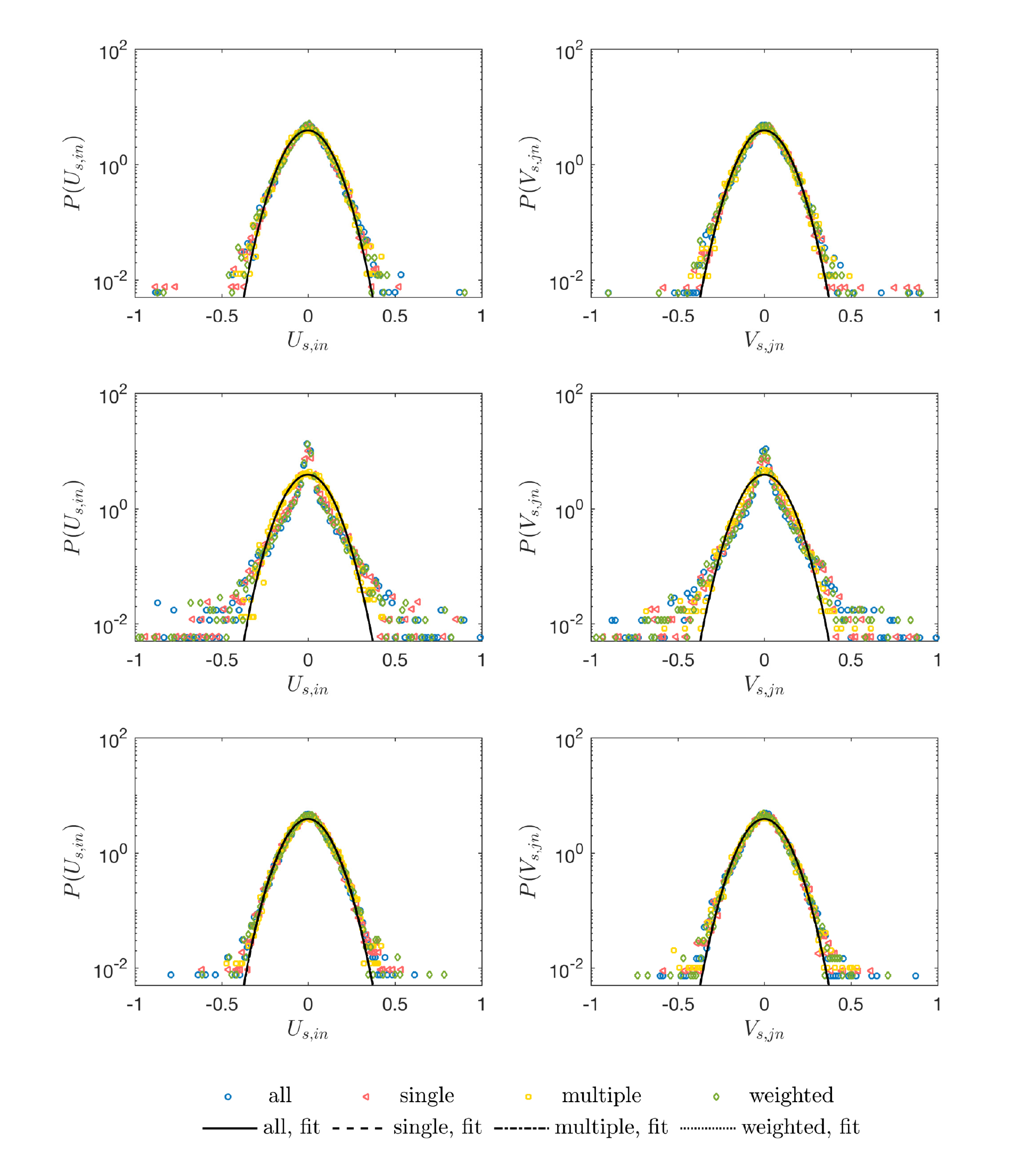}
    \caption{Probability density distributions of the entries of left and right singular vectors $U_{s,in}$ and $V_{s,jn}$ for the liquidity response across the whole market. Top: liquidity response to trade signs; middle: liquidity response to traded volumes; bottom: liquidity response to signed traded volumes. The fits to normal distributions are shown as black lines. In each subplot, the four cases of responses, \textit{i.e.}, the case of all trades, the case of single trades, the case of multiple trades and the case of weighted trades, are displayed.}
   \label{fig3.3}
   \end{center}
\end{figure*}

\subsection{Decomposing liquidity responses}
\label{sec3.3}

Figure~\ref{fig3.3} presents the probability density distributions of the entries $U_{s,in}$ or $V_{s,jn}$ of the singular vectors for liquidity responses across the whole market. The normal distributions are fitted for comparison. The mismatching between the empirical and the fitted distributions are visible at the heavy tails of the distributions. Thus, the connection between liquidity change and the latent factors cannot be coincidental, when the liquidity responds to trade signs, traded volumes or signed traded volumes. It is worth to keep in mind that the factors for liquidity may differ from the ones for prices. In the middle row of figure~\ref{fig3.3}, all cases except for the one of multiple trades present remarkable information in singular vectors, implying that the liquidity is highly sensitive to traded volumes.

\subsection{A unified description }
\label{sec3.4}

\begin{figure*}[pb]
  \begin{center}
    \includegraphics[width=0.9\textwidth]{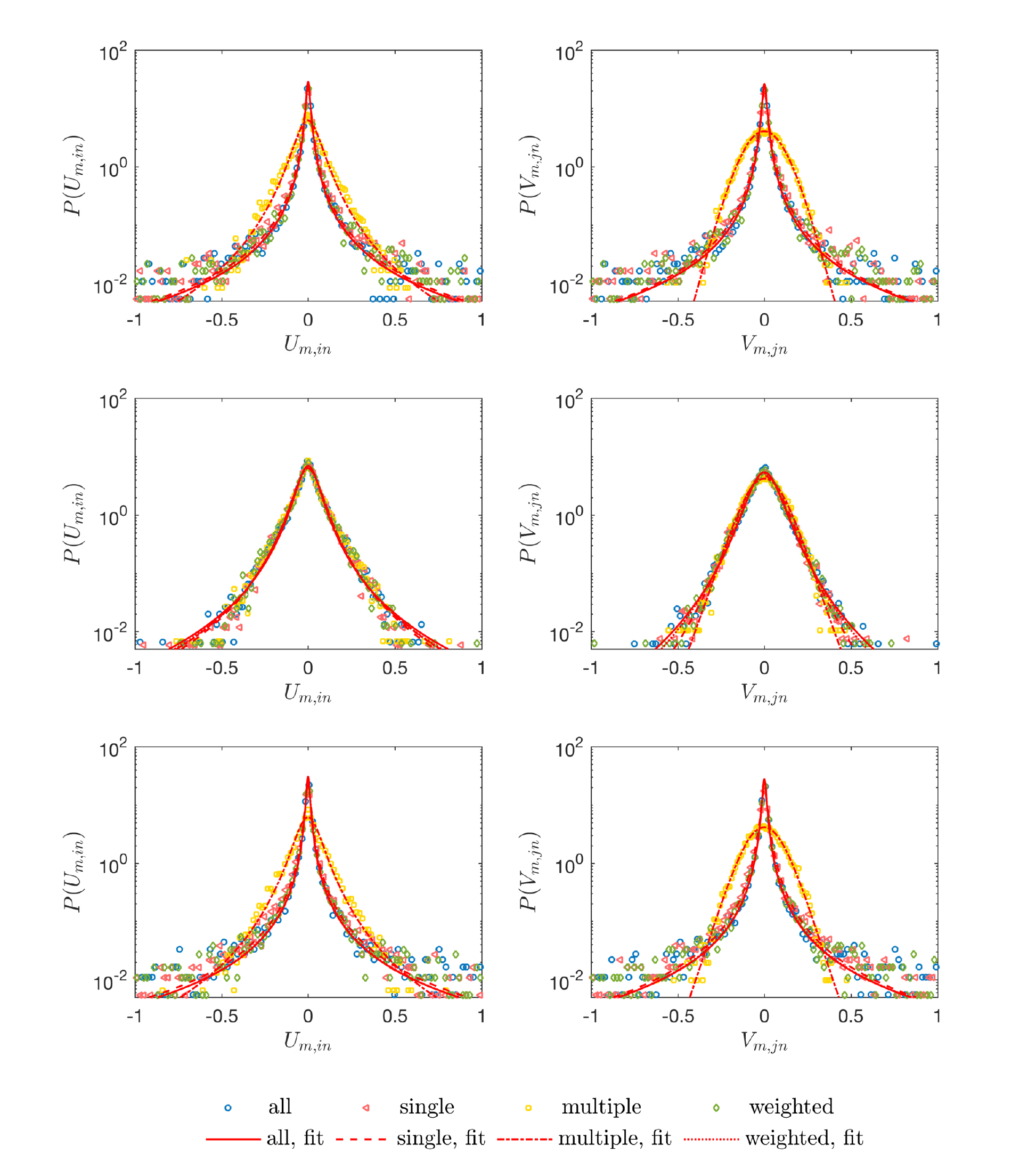}
    \caption{Probability density distributions of the entries of left and right singular vectors $U_{m,in}$ and $V_{m,jn}$ for the price response across the whole market. Top: price response to trade signs; middle: price response to traded volumes; bottom: price response to signed traded volumes. The fits to $t$ location-scale distributions are shown as red lines. In each subplot, the four cases of responses, \textit{i.e.}, the case of all trades, the case of single trades, the case of multiple trades and the case of weighted trades, are displayed.}
   \label{fig3.4}
   \end{center}
\end{figure*}

\begin{figure*}[pb]
  \begin{center}
    \includegraphics[width=0.9\textwidth]{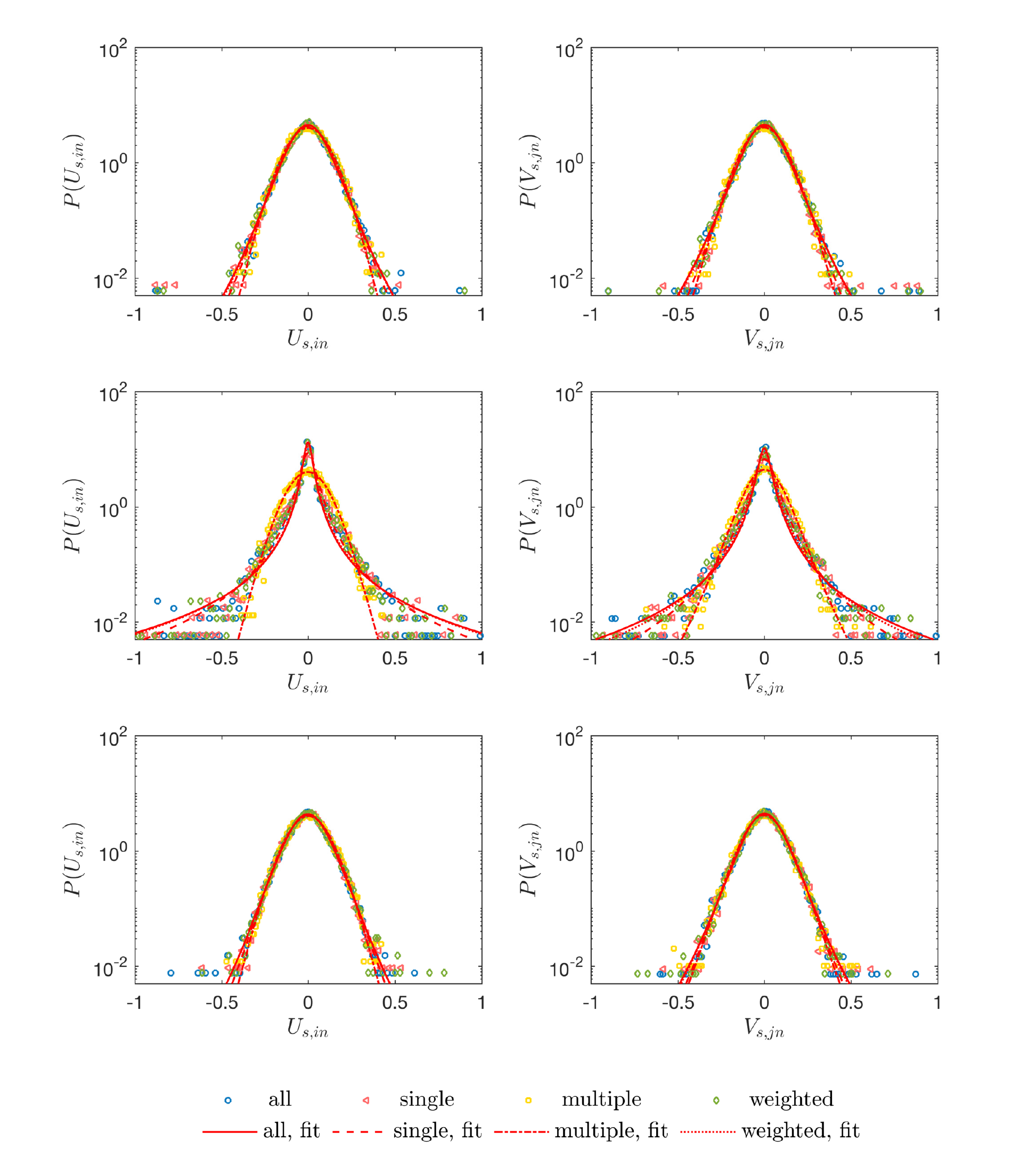}
    \caption{Probability density distributions of the entries of left and right singular vectors $U_{s,jn}$ and $V_{s,jn}$ for the liquidity response across the whole market. Top: liquidity response to trade signs; middle: spread response to traded volumes; bottom: spread response to signed traded volumes. The fits to $t$ location-scale distributions are shown as red lines. In each subplot, the four cases of responses, \textit{i.e.}, the case of all trades, the case of single trades, the case of multiple trades and the case of weighted trades, are displayed.}
   \label{fig3.5}
   \end{center}
\end{figure*}

The heavy tails are remarkable in the distributions of singular-vector entries. We introduce the $t$ location-scale distribution to quantify the potential information. The probability density function of the $t$ location-scale distribution, \textit{i.e.}, of the non-standardized Student's $t$-distribution~\cite{Jackman2009}, is given by
\begin{equation}
p(x)=\frac{\Gamma\left(\frac{\beta+1}{2}\right)}{\sigma\sqrt{\beta\pi} \Gamma \left(\frac{\beta}{2}\right)}\left[\frac{\beta+\left(\frac{x-\mu}{\sigma} \right)^2}{\beta} \right]^{-\left(\frac{\beta+1}{2} \right)} .
 \label{eq4.5}
\end{equation}
Here, $\Gamma(\cdot)$ is the gamma function, $\mu$ is the location parameter, $\sigma$ is the scale parameter, and $\beta$ is the shape parameter. When the shape parameter $\beta$ becomes very large, the distribution approaches the normal distribution. The smaller $\beta$, the heavier are the tails. Hence, by altering $\beta$, the $t$ location-scale distribution can either be a surrogate of the normal distribution or model the heavy-tailed distribution. This makes it appropriate for our purpose. For the price and the liquidity responses, the empirical result is fitted perfectly, especially for the heavy tails in the distribution, as shown in figures~\ref{fig3.4} and \ref{fig3.5}. The fitted values especially for $\beta$, listed in tables~\ref{tab3.1} and \ref{tab3.2}, corroborate nicely the findings stated in sections~\ref{sec3.2} and \ref{sec3.3}.

\begin{table*}[tbp]
\newcolumntype{C}{>{\arraybackslash}c@{\hskip0.035\textwidth}}
\newcolumntype{X}{>{\arraybackslash}c@{\hskip0.05\textwidth}}
\newcolumntype{R}{>{\arraybackslash}r@{\hskip 0.064\textwidth}}
\caption{Fit parameters for price responses}
\begin{center}
\begin{footnotesize}
\begin{tabular*}{\textwidth}{CXRRRcRRr}
\br
response &   cases of	&	\multicolumn{3}{c}{$U_{m,in}$} 	&&	\multicolumn{3}{c}{$V_{m,jn}$}	\\ \cline{3-5}  \cline{7-9}
to		&   trades		&	$\mu$	&	$\sigma$	&	$\beta$	&	&	$\mu$	&	$\sigma$	&	$\beta$	\\
\mr
\multirow{4}{*}{\makecell{trade \\signs}}	
		&	all		&	0.00021	&	0.011	&	0.980	&	&	0.00009	&	0.012	&	1.028	\\ 
		&	single	&	-0.00017	&	0.014	&	1.035	&	&	0.00012	&	0.017	&	1.131	\\ 
		&	multiple	&	-0.00190	&	0.057	&	2.304	&	&	-0.00127	&	0.097	&	21.156	\\
		&	weighted	&	0.00030	&	0.011	&	0.982	&	&	0.00010	&	0.012	&	1.032	\\ 
\mr
\multirow{4}{*}{\makecell{trade\\ volumes}}	
		&	all		&	0.00053	&	0.051	&	1.984	&	&	-0.00032	&	0.068	&	3.217	\\ 
		&	single	&	-0.00022 	&	0.057	&	2.357	&	&	-0.00177	&	0.081	&	5.307	\\ 
		&	multiple	&	0.00017	&	0.056	&	2.242	&	&	0.00119	&	0.093	&	11.273	\\
		&	weighted	&	-0.00118	&	0.054	&	2.151	&	&	-0.00161	&	0.072	&	3.678	\\ 
\mr
\multirow{4}{*}{\makecell{signed\\ trade \\volumes}} 
		&	all		&	-0.00019	&	0.010	&	0.957	&	&	-0.00033	&	0.011	&	1.008	\\ 
		&	single	&	0.00005	&	0.014	&	1.031	&	&	0.00009	&	0.017	&	1.139	\\ 
		&	multiple	&	-0.00024	&	0.058	&	2.362	&	&	-0.00125	&	0.094	&	12.794	\\ 
		&	weighted	&	0.00055	&	0.011	&	0.984	&	&	0.00044	&	0.012	&	1.034	\\
\br
\end{tabular*}
\end{footnotesize}
\end{center}
\label{tab3.1}
\end{table*}

\begin{table*}[htbp]

\newcolumntype{C}{>{\arraybackslash}c@{\hskip0.035\textwidth}}
\newcolumntype{X}{>{\arraybackslash}c@{\hskip0.05\textwidth}}
\newcolumntype{R}{>{\arraybackslash}r@{\hskip 0.062\textwidth}}
\caption{Fit parameters for liquidity responses}
\begin{center}
\begin{footnotesize}
\begin{tabular*}{\textwidth}{CXRRRcRRr}
\br
response &cases of	&	\multicolumn{3}{c}{$U_{s,in}$} 	&&	\multicolumn{3}{c}{$V_{s,jn}$}	\\ \cline{3-5}  \cline{7-9}
     to		&trades	&	$\mu$	&	$\sigma$	&	$\beta$		&&	$\mu$	&	$\sigma$	&	$\beta$	\\
\mr
\multirow{4}{*}{\makecell{trade \\signs}}
	&	all		&	-0.00136	&	0.085	&	6.491	&&	-0.00026	&	0.084	&	6.033	\\ 
	&	single	&	-0.00018	&	0.089	&	8.722	&&	0.00038	&	0.091	&	10.422	\\
	&	multiple	&	-0.00032	&	0.098	&	27.156	&&	-0.00226	&	0.097	&	18.985	\\ 
	&	weighted	&	0.00058	&	0.086	&	6.781	&&	0.00046	&	0.084	&	6.307	\\ 
\mr
\multirow{4}{*}{\makecell{trade\\ volumes}}
	&	all		&	-0.00035	&	0.024	&	1.073	&&	-0.00005	&	0.032	&	1.344	\\ 
	&	single	&	0.00092	&	0.039	&	1.500	&&	0.00091	&	0.053	&	2.148	\\ 
	&	multiple	&	-0.00129	&	0.098	&	23.204	&&	0.00088	&	0.087	&	7.223	\\ 
	&	weighted	&	0.00088	&	0.027	&	1.137	&&	0.00073	&	0.039	&	1.567	\\
\mr
\multirow{4}{*}{\makecell{signed\\ trade \\volumes}}
	&	all		&	-0.00132	&	0.088	&	7.774	&&	0.00054	&	0.085	&	6.549	\\ 
	&	single	&	0.00143	&	0.092	&	10.892	&&	0.00053 	&	0.091	&	9.371 	\\ 
	&	multiple	&	0.00105	&	0.097	&	21.939	&&	-0.00157	&	0.093	&	11.637	\\ 
	&	weighted	&	-0.00071	&	0.088	&	7.695 	&&	0.00017	&	0.087	&	7.044	\\
\br
\end{tabular*}
\end{footnotesize}
\end{center}
\label{tab3.2}
\end{table*}

\section{Relations between prices and liquidity}
\label{sec4}

Compared with the case of multiple trades, the case of single trades contains useful information. To study the relations between prices and liquidity, we focus on the case of single trades, which do not share their previous and following quotes with other trades. In section~\ref{sec4.1}, we define overlap matrices of factors. In section~\ref{sec4.2}, we analyze the structural characteristics with respect to overlap matrices in different cases. In section~\ref{sec4.3}, we dissect the overlap matrices by a singular value decomposition and discuss the relations between prices and liquidity.

\subsection{Overlap matrices}
\label{sec4.1}

To find out the common factors between price and liquidity changes, we introduce overlap matrices of factors. The overlap matrix~\cite{Benzaquen2017} is defined with the left singular vectors $\vec{U}_x$. We first normalize the entries $U_{x,in}$, $n=1,\ldots,N$, according to
\begin{equation}
\tilde{U}_{x,in}=\frac{U_{x,in}-\langle U_{x,in}\rangle_n}{\sigma(U_{x,in})_n} 
\label{eq4.1}
\end{equation}
where the average and the standard deviation for each stock $i$ are worked out over all $N$ factors. This defines the normalized left singular vectors $\tilde{\vec{U}}_{xi}$ and thereby the normalized matrix $\tilde{U}_x$, where $x$ stands for price changes when $x=m$ or liquidity changes when $x=s$. Hence, the $N\times N$ overlap matrix of factors reads
\begin{equation}
C_{ms} = \tilde{U}_m^T\tilde{U}_s\ .
\label{eq4.2}
\end{equation}
It measures the overlap of the $N$ factors of price change with the $N$ factors of liquidity changes. Likewise, the $N\times N$ overlap matrices for the factors of price changes and for the factors of liquidity changes are defined as
\begin{equation}
C_{mm} = \tilde{U}_m^T\tilde{U}_m  
\quad \mathrm{and} \quad
C_{ss} = \tilde{U}_s^T\tilde{U}_s \ ,
\label{eq4.3}
\end{equation}
respectively.

\subsection{Overlap structures}
\label{sec4.2}

\begin{figure*}[p]
  \begin{center}  
  \vspace*{-0.3cm}
    \includegraphics[width=0.77\textwidth]{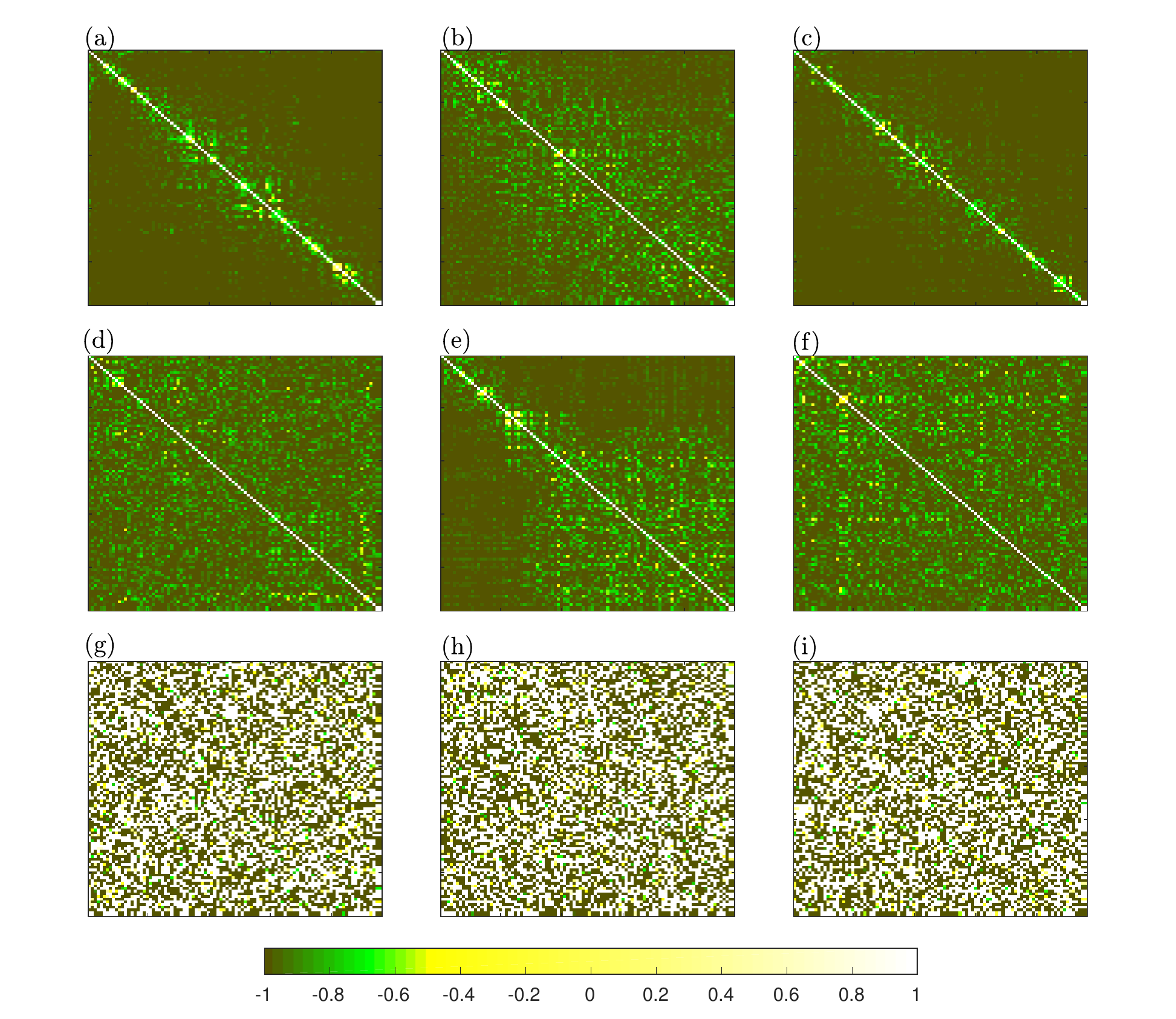}
    \vspace*{-0.3cm}
    \caption{Overlap matrices of factors $C_{mm}$ (top row), $C_{ss}$ (middle row) and $C_{ms}$ (bottom row), whose the factors are related to the trade signs (left column), the traded volumes (middle column) and the signed traded volumes (right column), respectively. In each subplot, the vertical axis is stock $i$, the horizontal axis is stock $j$.}
   \label{fig4.1}
   \end{center}
   \vspace*{-1cm}
\end{figure*}

\begin{figure*}[htbp]
  \begin{center}
    \includegraphics[width=0.77\textwidth]{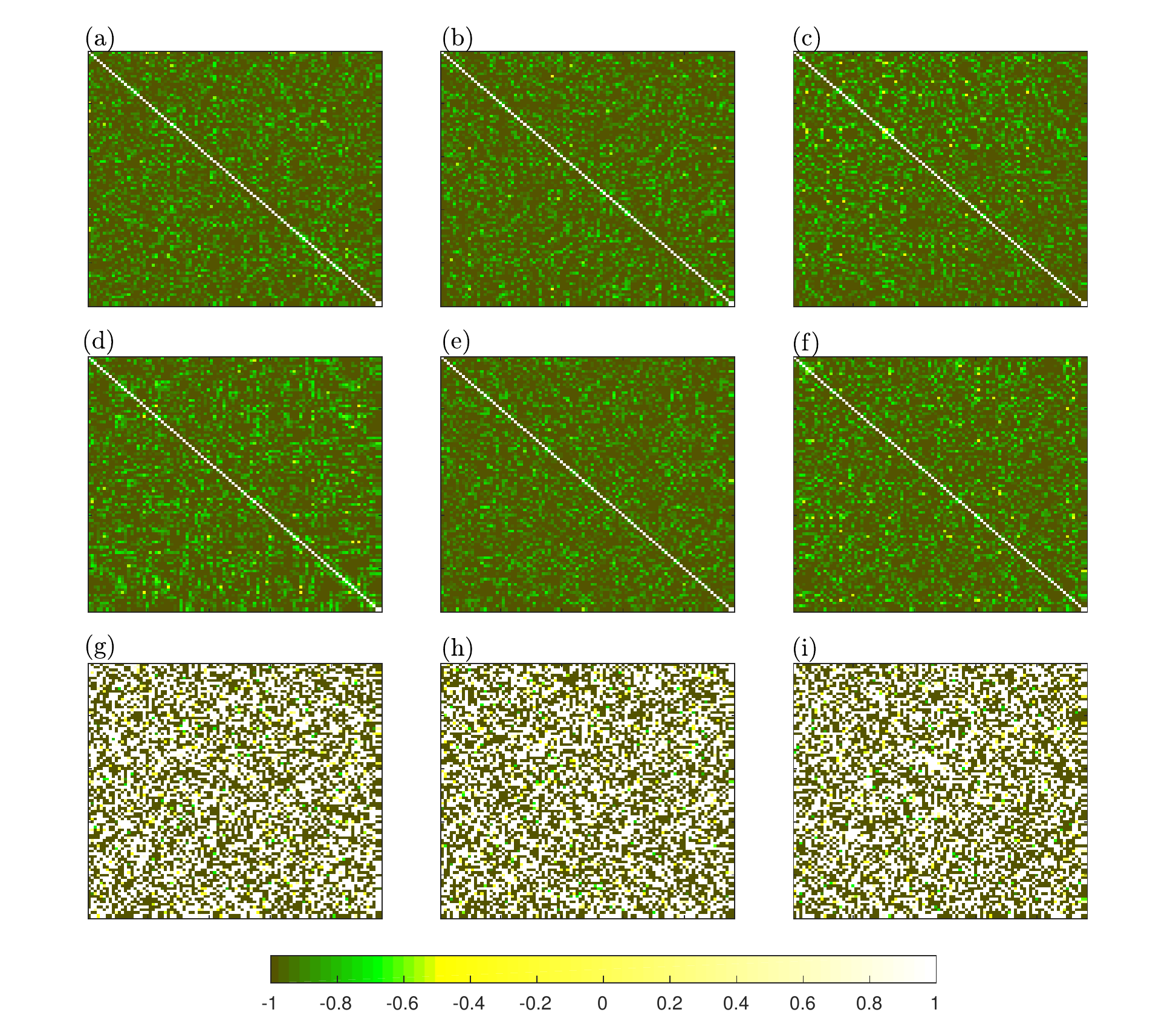}
    \vspace*{-0.3cm}
    \caption{Random overlap matrices, each subplot corresponds one-to-one to the subplot in figure~\ref{fig4.1}.}
   \label{fig4.2}
   \end{center}
    \vspace*{-0.5cm}
\end{figure*}

Figure~\ref{fig4.1} displays the overlap structure of factors, as measured with the matrices $C_{mm} $ (top row), $C_{ss}$ (middle row) and $C_{ms}$ (bottom row), where the factors are related to trade signs (left column), traded volumes (middle column) and signed traded volumes (right column), respectively. The overlaps are remarkable in the subplots $(a)$, $(b)$, $(c)$ and $(e)$. In particular, the factors of price changes are clearly related to the trade signs and the signed traded volumes. Thus, the price is easily moved by trade directions, \textit{i.e.}, buying or selling. On the other hand, the liquidity is significantly affected by the factors related to the traded volumes rather than by others. These overlapping features in $C_{mm} $ and $C_{ss}$ are striking when comparing with the features in random overlap matrices, shown in figure~\ref{fig4.2}. The random overlap matrices result from the random response matrices in which the mean values and standard deviations are the same as the empirical response matrices. In figure~\ref{fig4.2}, the small overlaps are randomly distributed in each random overlap matrix, quite different from the empirical cases. 

The overlap matrix visualizes the overlapping of factors that individually change the price or the liquidity, but it fails to identify the overlapping of factors that jointly drive the price and the liquidity, because the largely positive and negative overlaps are mixed and look like the random patterns in the subplots $(g)$---$(i)$ of figure~\ref{fig4.1}. Therefore, the singular value decomposition is applied once more to the overlap matrices.


\subsection{Decomposing overlap matrices}
\label{sec4.3}

\begin{figure*}[p]
  \begin{center}
  \vspace*{-0.3cm}
    \includegraphics[width=0.74\textwidth]{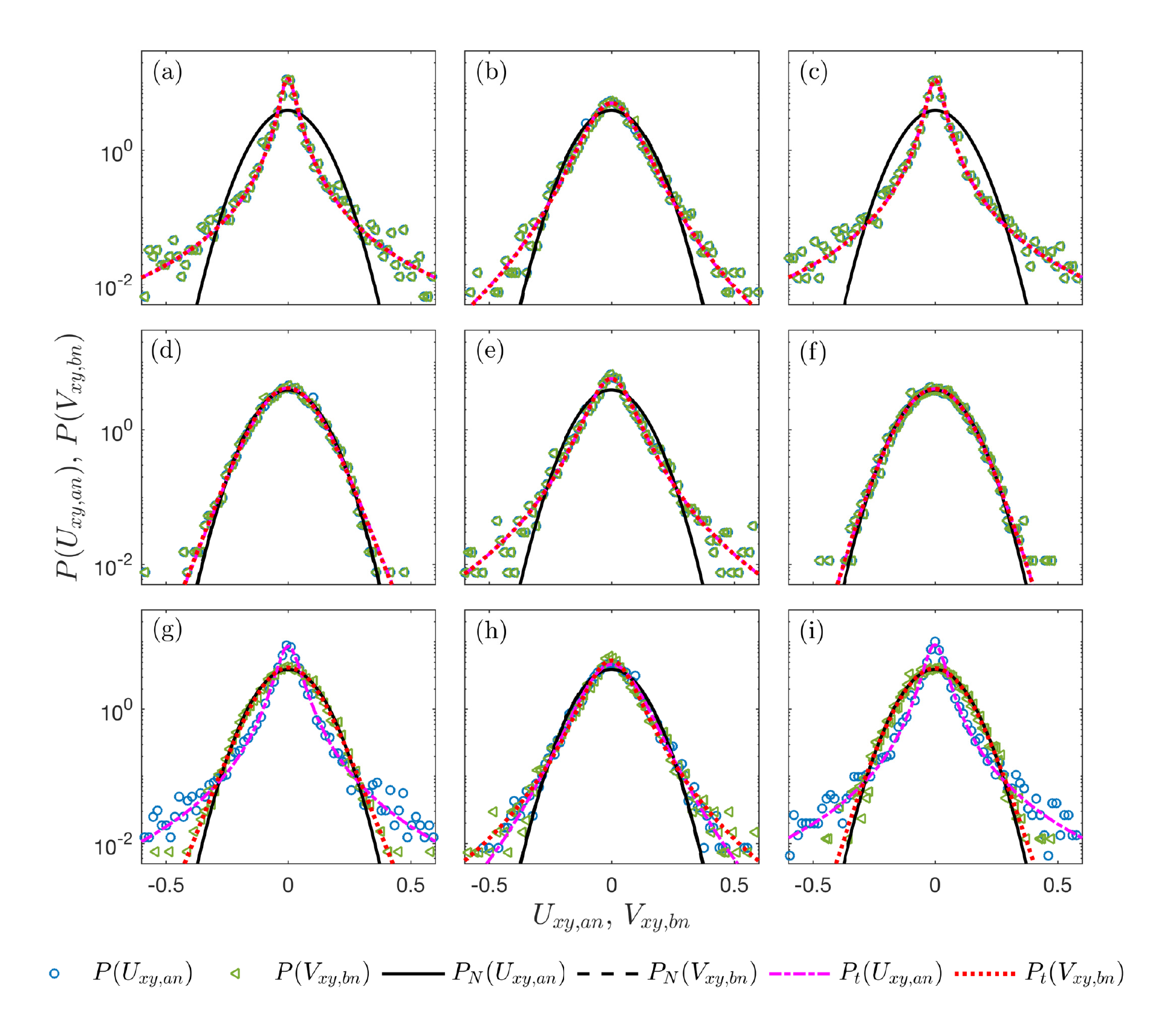}
    \vspace*{-0.5cm}
    \caption{Probability density distributions of entries $U_{xy,an}$ and $V_{xy,bn}$ of singular vectors for the overlap matrices $C_{mm} $ (top row), $C_{ss} $ (middle row) and $C_{ms}$ (bottom row). The factors are related to the trade signs (left column), the traded volumes (middle column) and the signed traded volumes (right column), respectively. All empirical distributions are fitted by normal distributions $P_N(U_{xy,an})$, $P_N(V_{xy,bn})$ and by $t$ location-scale distributions $P_t(U_{xy,an})$, $P_t(V_{xy,bn})$.}
   \label{fig4.3}
   \end{center}
    \vspace*{-1cm}
\end{figure*}
\begin{figure*}[htbp]
  \begin{center}
      \includegraphics[width=0.74\textwidth]{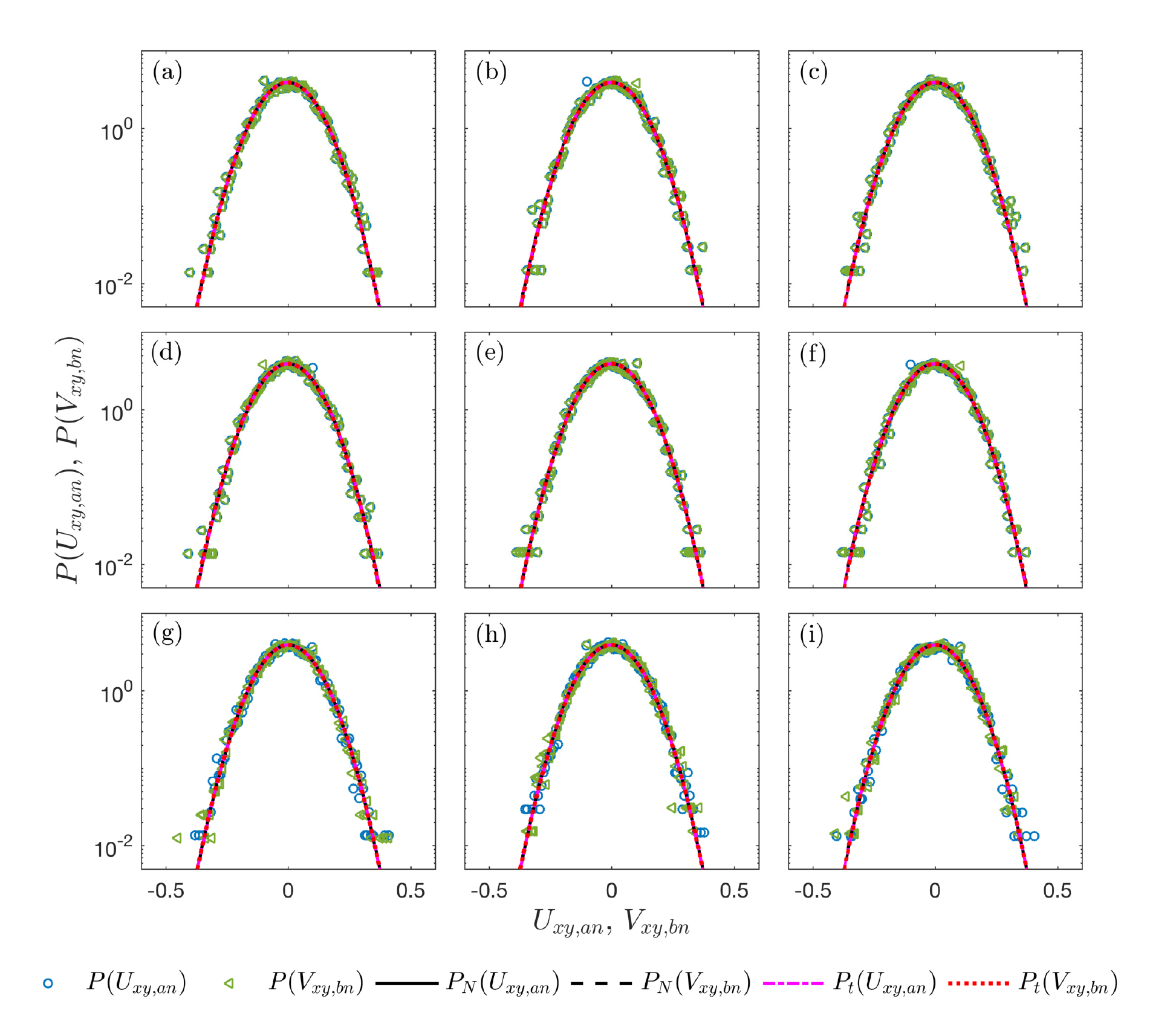}
    \vspace*{-0.5cm}
    \caption{The probability density distributions of entries $U_{xy,an}$ and $V_{xy,bn}$ of singular vectors for the random overlap matrices, where each subplot one-to-one corresponds to the subplot in figure~\ref{fig4.3}. All distributions are fitted by normal distributions $P_N(U_{xy,an})$, $P_N(V_{xy,bn})$ and by $t$ location-scale distributions $P_t(U_{xy,an})$, $P_t(V_{xy,bn})$.}
   \label{fig4.4}
   \end{center}
    \vspace*{-1cm}
\end{figure*}

For a given $x$ and $y$, the overlap matrix $C_{xy}$ is decomposed into left and right singular vectors which are the columns of orthogonal $N\times N$ matrices $U_{xy}$ and $V_{xy}$, respectively. The corresponding singular values are ordered in the diagonal matrix $S_{xy}$. Thus, the decomposition reads,
\begin{equation}
C_{xy}=U_{xy}S_{xy}V_{xy}^T \ .
\label{eq4.4}
\end{equation}
The entries $U_{xy,an}$ and $V_{xy,bn}$ with $a,b, n=1,\ldots,N$ in the singular vector matrices measure the correlation between the $n$-th common factors and the $a$-th individual factors of $x$ and the correlation between the $n$-th common factors and the $b$-th individual factors of $y$, respectively. In figure~\ref{fig4.3}, we show the probability densities of the left and right singular vectors of overlap matrices $C_{mm}$, $C_{ss}$ and $C_{ms}$, where both the normal distribution and the $t$ location-scale distribution are fitted to the empirical distributions. Table~\ref{tab4.1} lists all fit parameters. For comparison, the same procedure is carried out for the random overlap matrices, shown in figure~\ref{fig4.4}. 

In figure~\ref{fig4.3}, the subplots $(a)$, $(b)$, $(c)$ and $(e)$ show the heavy tails for $C_{mm}$ and $C_{ss}$, coinciding with the larger overlaps in figure~\ref{fig4.1}. In particular, the irregular patterns for $C_{ms}$ in figure~\ref{fig4.1} are verified to be non-random. The information encoded in the overlap matrices are quantified by the $t$ location-scale distribution. Among the trade signs, the traded volumes and the signed traded volumes, the factors related to the traded volumes are identified to significantly interconnect the price change with the liquidity change. Put differently, the traded volume plays an important role when the price is changed by the liquidity. This is plausible, as the trades from the stock itself are able to affect the short-run liquidity by eating up the volumes in the best quote, and move the price to another level immediately. In contrast, the factors related to the trade signs and the signed traded volumes generate much weak interconnections between price changes and liquidity changes. This reveals that the liquidity changes are almost cancelled out by the opposite trade directions. However, we still cannot ignore the factors that lead to the weak liquidity responses to the trade signs and the signed traded volumes, as the resulting liquidity imbalance contributes to the price change.

\begin{table*}[tbp]
\newcolumntype{C}{>{\arraybackslash}c@{\hskip0.03\textwidth}}
\newcolumntype{R}{>{\arraybackslash}r@{\hskip 0.042\textwidth}}
\caption{Fit parameters for correlations of factors}
\begin{center}
\begin{footnotesize}
\begin{tabular*}{1\textwidth}{CCRRrcRRr}
\br
correlation & factors &	\multicolumn{3}{c}{$U_{xy,an}$} 	&&	\multicolumn{3}{c}{$V_{xy,bn}$}	\\  \cline{3-5}  \cline{7-9}
between	  & related to&	$\mu$	&	$\sigma$	&	$\beta$		&&	$\mu$	&	$\sigma$	&	$\beta$	\\
\hline
\multirow{3}{*}{\makecell{factors of\\ price change}}
	&\makecell{trade signs}		&	0.00003	&	0.029	&	1.378	&&	0.00003	&	0.029	&	1.378	\\  
	&\makecell{traded volumes}	&	-0.00183	&	0.075	&	4.107	&&	0.00070	&	0.075	&	4.109	\\ 
	&\makecell{signed volumes}	&	0.00062 	&	0.030	&	1.401	&&	 0.00062	&	0.030	&	1.401	\\
\hline
\multirow{3}{*}{\makecell{factors of \\  liquidity change}}
	&\makecell{trade signs}		&	0.00151	&	0.095	&	14.370	&&	-0.00087	&	0.095	&	14.472	\\ 
	&\makecell{traded volumes}	&	-0.00128	&	0.064	&	2.889	&&	-0.00128	&	0.064	&	2.889	\\ 
	&\makecell{signed volumes}	&	-0.00117	&	0.098	&	26.769	&&	-0.00117	&	0.098	&	26.769	\\
\hline
\multirow{3}{*}{\makecell{ factors of price\\ change and of \\ liquidity change}}
	&\makecell{trade signs}		&	0.00108	&	0.041	&	1.784	&&	0.00148	&	0.094	&	13.893	\\ 
	&\makecell{traded volumes}	&	0.00141	&	0.083	&	5.655	&&	-0.00112	&	0.070	&	3.421	\\ 
	&\makecell{signed volumes}	&	-0.00063	&	0.038	&	1.630	&&	-0.00129	&	0.097 	&	22.677	\\
\br
\end{tabular*}
\end{footnotesize}
\end{center}
\label{tab4.1}
\end{table*}

\section{Conclusions} 
\label{sec5}
 
We explored the whole market response to trade signs, traded volumes and signed traded volumes, focusing on the primary price and liquidity responses. Utilizing the singular value decomposition, the response matrices were dissected into the left and right singular vectors and the corresponding singular values. We analyzed the statistics properties of the singular vectors, where the left singular vectors correlate the price or liquidity change with the latent factors, and the right singular vectors correlate these factors with the trading information. In our study, the trade information includes the trade signs, the traded volumes and the signed traded volumes. We found that the heavy-tailed distributions of singular vectors either for price responses or for liquidity responses are well described by $t$ location-scale distributions. The price responds significantly to the trade signs and signed traded volumes, whereas the liquidity is very sensitive to the traded volumes.

We also looked at the overlap matrices for the factors that individually change the price or the liquidity and for the factors that joint change the price and the liquidity. The overlap matrices reveal non-random structures. The overlaps are remarkable when the factors of price changes are related to the three kinds of trading information and when the factors of liquidity changes are related to the traded volumes. By carrying out a singular value decomposition for the overlap matrices, we found that the factors related to the traded volumes interconnect significantly the price change with the liquidity change. Hence, the unsigned traded volumes appear critical for the price change caused by the liquidity. On the other hand, the bid-ask spread can be enlarged either by a buy trade or by a sell trade, resulting in a reduction of liquidity. If the market is efficient, the liquidity responses to trade signs or signed traded volumes should be counterbalanced. However, we found the two kinds of liquidity responses are weak but cannot be ignored, as they imply an imbalance of liquidity, which is related to the price change.

\begin{appendices}

\section{Stock information}
\label{appA}

Table~\ref{appA.1} lists 96 stocks used in this study and their average daily numbers of trades. The daily number of trades is restricted to the intraday trading time from 9:40 to 15:50 EST, and the average is performed over five trading days from March 7th to March 11th in 2016. Here, the information of trades is obtained by reconstructing the order book with the TotalView-ITCH data set.

\renewcommand{\thetable}{A.\arabic{table}}  
\setcounter{table}{0} 
\begin{table*}[tbp]
\newcolumntype{L}{>{\arraybackslash}l@{\hskip0.02\textwidth}}
\newcolumntype{R}{>{\arraybackslash}r@{\hskip 0.1\textwidth}}
\caption{Averaged daily number of trades over five trading days}
\begin{center}
\begin{footnotesize}
\begin{tabular*}{\textwidth}{LRLRLRLR}
\br
stock & number & stock & number & stock & number & stock & number \\
\mr
AAL		&	4563	&	COST	&	3487	&	JD		&	3596	&	REGN	&	1300	\\ 
AAPL	&	13598&	CSCO	&	3273	&	KHC		&	3140	&	ROST	&	3868	\\ 
ADBE	&	4553	&	CTSH	&	4823	&	KLAC	&	1803	&	SBAC	&	1935	\\ 
ADI		&	2931	&	CTXS	&	2477	&	LBTYA	&	2759	&	SBUX	&	5719	\\ 
ADP		&	2954	&	DISCA	&	3152	&	LLTC	&	2300	&	SIRI		&	514	\\ 
ADSK	&	3389	&	DISH	&	2261	&	LMCA	&	1585	&	SNDK	&	3687	\\ 
AKAM	&	2439	&	DLTR	&	4021	&	LRCX	&	3826	&	SPLS	&	1108	\\ 
ALXN	&	2466	&	EA		&	4708	&	LVNTA	&	1063	&	SRCL	&	1588	\\ 
AMAT	&	2066	&	EBAY	&	2850	&	MAR		&	3495	&	STX		&	4056	\\ 
AMGN	&	5132	&	EQIX	&	1615	&	MAT		&	2918	&	SYMC	&	1784	\\ 
AMZN	&	5376	&	ESRX	&	6144	&	MDLZ	&	3666	&	TRIP		&	3473	\\ 
ATVI		&	3882	&	EXPD	&	2310	&	MNST	&	1591	&	TSCO	&	1535	\\ 
AVGO	&	5518	&	FAST	&	2816	&	MSFT	&	9245	&	TSLA	&	3367	\\
BBBY	&	2590	&	FB		&	14921&	MU		&	2351	&	TXN		&	3479	\\ 
BIDU	&	2729	&	FISV		&	1856	&	MYL		&	5969	&	VIAB		&	3769	\\ 
BIIB		&	2818	&	FOXA	&	2388	&	NFLX	&	9164	&	VIP		&	217	\\ 
BMRN	&	2135	&	GILD		&	11681&	NTAP	&	2210	&	VOD		&	926	\\ 
CA		&	1531	&	GOOG	&	4426	&	NVDA	&	2935	&	VRSK	&	1264	\\ 
CELG	&	6742	&	GRMN	&	1909	&	NXPI		&	3824	&	VRTX	&	3037	\\ 
CERN	&	3440	&	HSIC	&	674	&	ORLY	&	1837	&	WDC	&	6662	\\ 
CHKP	&	2030	&	ILMN	&	1860	&	PAYX	&	1838	&	WFM	&	3775	\\ 
CHRW	&	2021	&	INTC		&	3933	&	PCAR	&	3315	&	WYNN	&	4046	\\ 
CHTR	&	2650	&	INTU		&	2299	&	PCLN	&	1029	&	XLNX	&	2450	\\ 
CMCSA	&	5984	&	ISRG	&	616	&	QCOM	&	7030	&	YHOO	&	6258	\\ 
\br
\end{tabular*}
\end{footnotesize}
\end{center}
\label{appA.1}
\end{table*}

\end{appendices}

\section*{References}
\addcontentsline{toc}{section}{References}

\providecommand{\newblock}{}

\end{document}